\documentclass[submission]{eptcs}
\pdfoutput=1

\usepackage{graphicx}
\usepackage{amsthm}
\usepackage{amsmath}
\usepackage{amsfonts}
\usepackage{proof}
\usepackage{latexsym}
\usepackage{paralist}
\usepackage{stmaryrd}

\newtheoremstyle{example}{\topsep}{\topsep}%
     {}
     {}
     {\bfseries}
     {.}
     { }
     {\thmname{#1}\thmnumber{ #2}\thmnote{ (#3)}}

\theoremstyle{example}

\newtheorem{definition}{Definition}

\newcommand{\SB}[1]{\llbracket #1 \rrbracket}

\newcommand{\ent}{\mathrel{{:}{-}}}

\newcommand{\used}{\mathrel{\mathsf{used}}}
\newcommand{\wgb}{\mathrel{\mathsf{wasGeneratedBy}}}
\newcommand{\wtb}{\mathrel{\mathsf{wasTriggeredBy}}}
\newcommand{\wdf}{\mathrel{\mathsf{wasDerivedFrom}}}

\newcommand{\predicts}{\rightsquigarrow}

\newcommand{\prov}[2][]{\mathsf{P}_{#1}#2}

\title{Causality and the Semantics of Provenance}
\author{James Cheney
\institute{LFCS\\University of Edinburgh}
\email{jcheney@inf.ed.ac.uk}
}
\def\titlerunning{}
\def\authorrunning{J. Cheney}
\begin{document}
\maketitle

\begin{abstract}
  Provenance, or information about the sources, derivation, custody or
  history of data, has been studied recently in a number of contexts,
  including databases, scientific workflows and the Semantic Web.
  Many provenance \emph{mechanisms} have been developed, motivated by
  informal notions such as influence, dependence, explanation and
  causality.  However, there has been little study of whether these
  mechanisms formally satisfy appropriate \emph{policies} or even how
  to formalize relevant motivating concepts such as causality.  We
  contend that mathematical models of these concepts are needed to
  justify and compare provenance techniques.  In this paper we review
  a theory of causality based on \emph{structural models} that has
  been developed in artificial intelligence, and describe work in
  progress on a causal semantics for provenance graphs.
\end{abstract}

\section{Introduction}

Provenance is a general term referring to the origin, history, chain
of custody, derivation or process that yielded an object.  In analog
settings such as art and archaeology, such information is essential
for understanding whether an artifact is authentic, valuable or
meaningful.  In the digital world, provenance is now recognized as an
important problem because it is very easy to silently alter or forge
digital information.  We already pay a price because of the lack of
robust mechanisms for recording and managing provenance: serious
economic losses have been incurred due to the lack of provenance
on the Web~\cite{wsj}, and lack of transparency of scientific
processes and results is routinely used to sow confusion and doubt
about climate change~\cite{climategate}.

Much work on provenance considers the following basic scenario: we
have some input data and some complex process that will be run on the
input, for example a large program, possibly split into many smaller
jobs and executed in parallel or on a distributed system.  In this
setting, a proposed solution generally records some additional
information as the program runs.  The additional information is often
called provenance and it is supposed to provide an ``explanation''
showing how the results were obtained.

Of course, this is a loose specification if we do not clarify what we
mean by an explanation (apart from whatever provenance information
happens to be recorded by the system).  There are at least two
obvious-seeming choices, neither of which seems satisfactory in
practice:
\begin{itemize}\item 
  First, we might record the program that was run along with its input
  data (and any intermediate inputs such as user or network
  interactions).  This, at least, allows us to rerun the program later
  and check that we get the same result, and it also allows us to vary
  the inputs to see how changes affect the output.  But this is nearly
  useless as an explanation, especially for end-users who are not (and
  should not be expected to be) proficient at debugging black-box
  systems.  Moreover, the inputs and outputs may be huge (for example,
  gigabytes of climate data), and it may not be possible for users to
  manually inspect the data.  In the longer term, just recording the
  program is also problematic since the computational environment in
  which the program runs will change --- in some sense, this is also
  an ``input'' that we cannot feasibly record.
\item Second, we might record everything that can be recorded
  about the computation, in the hope that it might someday be useful.
  This also sounds straightforward but is surprisingly difficult to
  pin down, since ``everything that can be recorded'' can be
  interpreted in many different ways.  Should we record every function
  call?  Every instruction? Every molecular interaction?  Do we need
  to record what the programmer had for breakfast on the day the
  program was written?  Clearly we have to stop somewhere, and for
  efficiency reasons we should probably stop far short of any
  reasonable definition of ``everything''.
\end{itemize}
Most extant approaches pick some intermediate point between these two
extremes, committing to some (often not explicitly stated) choices
about what is important about the computation that should be recorded
in its provenance.

For example, in databases, there are models such as
where-provenance~\cite{buneman01icdt} (tracking the ``sources'' of copied
data), lineage~\cite{cui00tods}, why-provenance~\cite{buneman01icdt}
or how-provenance~\cite{green07pods} (tracking tuples ``used by'' or
that ``justify'' a result tuple), or dependency
provenance~\cite{cheney07dbpl} (a computable approximation of the
information flow behavior of the program).

Provenance has also been studied extensively in other settings,
particularly ``scientific workflow'' systems
(e.g.~\cite{taverna,kepler,petri}).  Scientific workflows are usually
high-level, visual programming languages, often based on dataflow or
Petri-net models of concurrent computation, and often executed on grid
or cloud computing platforms.  This architecture has the advantage
that it puts considerable computational power into the hands of
scientists without forcing them to learn how to program parallel or
distributed systems at a low level in C++ or Java.  However, it also
has a serious drawback: distributing a program over a heterogeneous
network dramatically increases the number of things that can go wrong,
typically makes the computation nondeterminstic and makes it hard for
the user to trust the results.  

Scientists are reluctant to publish results based on programs that may
contain subtle bugs, and whose behavior is different every time they
are run, or depends on libraries or other environmental factors in
subtle ways.  Provenance is perceived as important for helping users
understand whether results of such computations are repeatable and
trustworthy, and in particular for scientists to be able to judge the
scientific validity of results they may wish to publish.

The work on database provenance is distinctive in that several
different formal models have now been defined for database query
languages with well-understood semantics.  This makes it easier to
compare, relate and generalize these approaches, though such
comparisons are only starting to
appear~\cite{cheney09ftdb,green07pods}.  For most of these models,
there are semantic guarantees (or even exact semantic
characterizations) relating the provenance records to the denotation
of the program.  On the other hand, for workflow provenance, formal
definitions of the meaning of workflow programs have only started to
appear recently (see for
example~\cite{Sroka2009,DBLP:conf/dils/HiddersKSTB07}), while the
provenance semantics of these tools is usually specified informally,
at best~\cite{taverna}.  As a result there is a confusing variety of
models and styles of provenance for workflows.

To address this problem, there has been an ongoing community effort,
centered on a series of ``Provenance
Challenges''~\cite{prov-challenge}, to understand and compare the
qualitative behavior of these different systems and synthesize a
common format for exchanging provenance among them.  This effort has
recently yielded a draft Open Provenance Model, or OPM~\cite{opm11}.
Instances of this model are graphs whose nodes represent agents,
processes or artifacts and whose edges represent dependence,
generation or control relationships.  The OPM has ``semantics'' in the
sense of the Semantic Web, in that the nodes and edges are expected to
have names that are meaningful to reasonably well-informed users.
The
OPM standard draws heavily on informal motivations such as
``provenance is the process that led to a result'' and ``edges denote
causal relationships linking the cause to the effect''.  But while the
OPM specifies a graph notation, controlled vocabulary for the edges,
and inference rules for inferring new edges from existing edges, it
does not have a ``semantics'' in the denotational or operational sense
by which we might judge whether a graph is consistent or complete or
whether inferences on the graph are valid.

In this paper, we investigate the use of \emph{structural causal
  models}~\cite{pearl00causal} as a semantics for these graphs, and
relate the informal motivations invoked in defining OPM graphs with
the formal definitions of \emph{actual cause} and \emph{explanation}
due to Halpern and Pearl~\cite{halpern05bjps-1,halpern05bjps-2}.  We
do not argue that structural causal models provide the only or best
causal account of provenance.  However, structural causal models are
quite close to OPM-style provenance graphs (modulo cosmetic
differences), so the analogy is compelling.  Moreover, structural
models have been studied extensively
(e.g. ~\cite{pearl00causal,halpern05bjps-1,halpern05bjps-2,eiter02ai,eiter04ai,eiter06ai})
and have proven useful to both philosophical accounts of scientific
explanation~\cite{woodward} and psychological theories of
understanding~\cite{sloman05causal}.  Nevertheless, it may be
enlightening to apply other mathematical theories of causality and
explanation to provenance, or investigate variations and extensions of
Halpern and Pearl's approach.

The broader aim of this paper is to argue by example that semantics
(in the mathematical sense) is badly needed for research on
provenance.  One of the major motivations for provenance is to improve
scientific communication, by allowing scientists to generate and
exchange computational ``explanations'' or ``justifications'' of their
results.  In biology, for example, some journals now require both data
and workflow programs describing how results were obtained, and some
scientists anticipate that scientific publication will evolve into
richer documents incorporating text, data, and
computation~\cite{adventures}.  However, if the techniques used to do
so are poorly specified and unverified then we can expect errors and
confusion.  Programming languages and semantics researchers can and
should be involved in making sure that these techniques are clearly
described and robust, to help ensure that scientific communications
retain long-term value as they gain computational structure.

\section{Examples}

Before delving into technical details, we give a high-level example
comparing OPM-style provenance graphs with structural causal models.

\begin{figure}
  \begin{center}
    \includegraphics[scale=0.25]{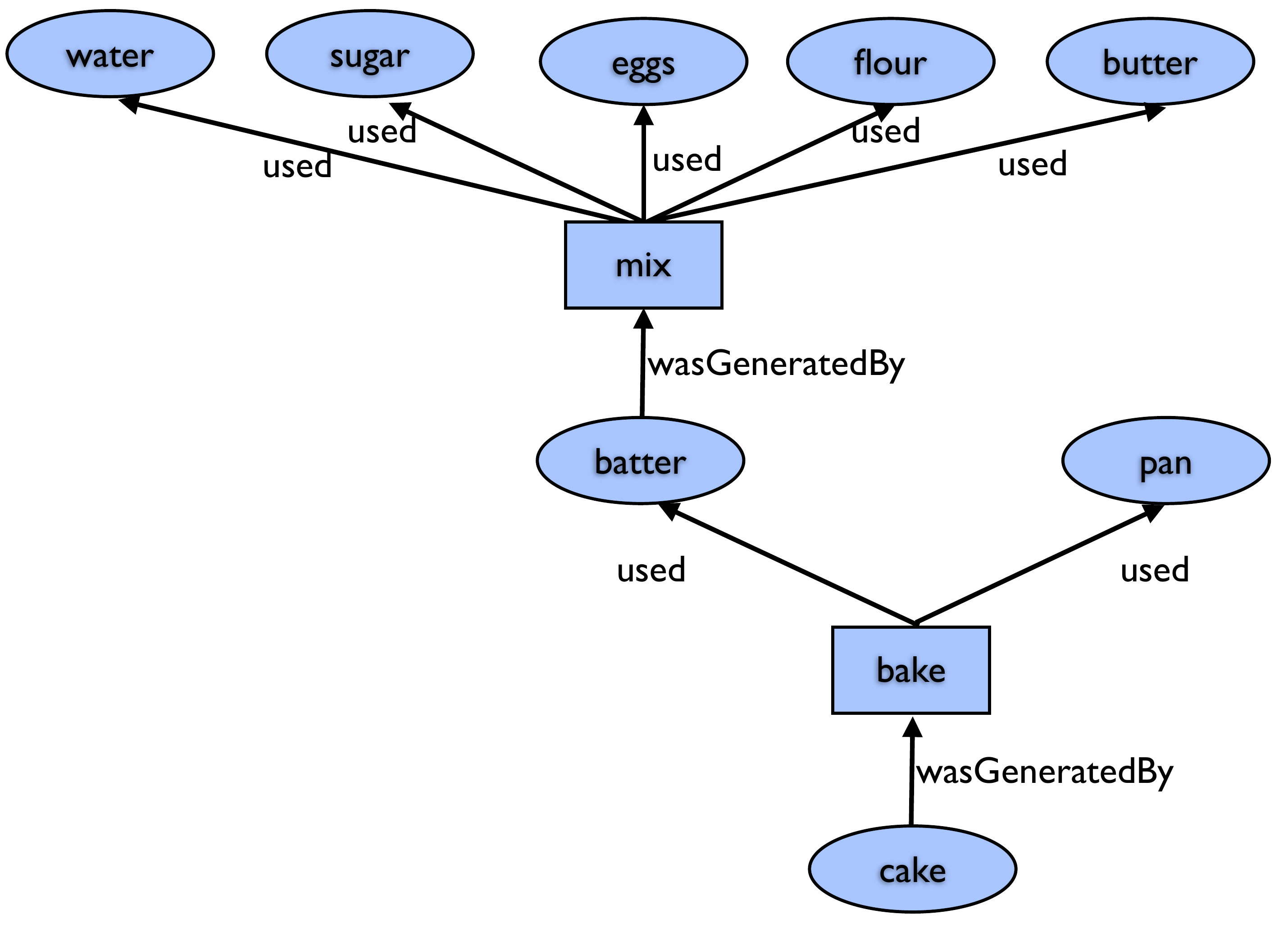}
    \qquad
    \includegraphics[scale=0.25]{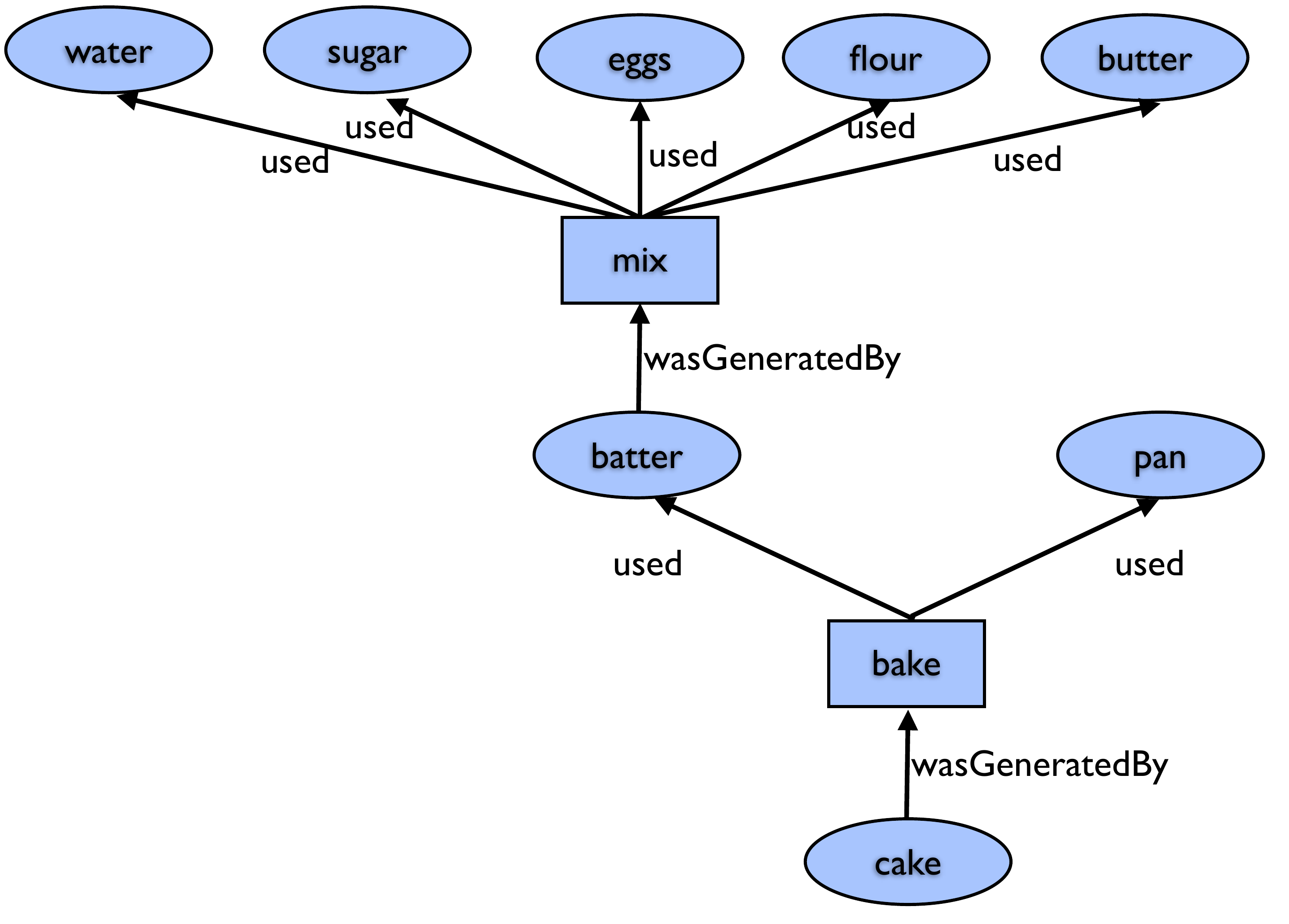}
\begin{eqnarray*}
  Mix &:=& (Water \wedge Sugar \wedge Eggs \wedge Flour \wedge
  Butter) \oplus U_1\\
  Batter &:=& Mix \oplus U_2\\
  Bake &:=& (Batter\wedge Pan) \oplus U_3\\
  Cake &:=& Bake \oplus U_4
\end{eqnarray*}
  \end{center}
  \caption{Top left: An OPM graph.  Top right: the corresponding
    causal model.  Bottom: a straight-line code description of both
    processes.  Variables $U_1,U_2,U_3,U_4$ are exogenous variables
    (error terms, inputs) abstracting away unknown environmental
    influences that are not explicitly modeled.}
\label{fig:opm-causal}
\end{figure}

The left-hand side of Figure~\ref{fig:opm-causal} shows a simple OPM
graph, based on a standard example showing the ``provenance of a
cake'' ~\cite{opm11}.  The right-hand side shows a structural causal
model, depicted as a graph.  These two graphs are intentionally very
similar.  In the OPM graph, the ovals denote ``artifacts'' (flour,
water, pan, cake) while the boxes denote ``processes'' (baking,
mixing).

The OPM standard does not specify a semantics of provenance graphs as
computations that might take place in the future; rather, it specifies
a syntax for graphs that are supposed to describe processes that have
taken place in the past.  Nevertheless, it is natural to read an OPM
graph as an expression or straight-line program that can be rerun to
produce the output in the same way as the original process did.

Conversely, a causal model such as the one in Figure
\ref{fig:opm-causal} does have an intended computational semantics.
Each node in a causal model is equipped with a function specifying how
to compute the value of the node given values for the parent nodes.
An acyclic causal graph can be viewed as a piece of straight-line
code, assigning values to variables as shown along the bottom of
Figure~\ref{fig:opm-causal}.  Here, we considered a simplistic
boolean-valued model whose transfer functions are conjunctions.  We
can interpret these definitions as saying that if all of the needed
ingredients for each step of the process are present, then that step
will succeed.

Causal models also typically distinguish between the set of
\emph{exogenous} parameters $U$ and \emph{endogenous} variables $V$.
The former are meant to represent unknowns, measurement error, or
environmental factors not otherwise taken into account in the model.
Here, for example, we included one exogenous parameter $U_i$ for each
computation step, representing the possibility that something might go
wrong.

Causal models are closely related to Bayesian networks, and causal
models can also be given a probabilistic interpretation.  For example,
to model a scenario where the oven explodes and destroys the cake with
probability 0.01, we can employ a probability distribution giving
$U_3$ the expected value 0.01.

In artificial intelligence, probabilistic causal models are used
either to represent domain knowledge that enables a system to reason
about cause and effect, or as a representation of hypotheses about
some data whose behavior is believed to be causal.  Bayesian inference
can then be used to learn causal models from data. This kind of
inference is appropriate for data generated by controlled experiments
where a single parameter can be varied to identify cause-effect
relationships, but can also be used to analyze situations with less
ideal experimental designs~\cite{pearl00causal}.

In this paper, however, we will consider only the deterministic form
of causal models and focus on their use as a semantic tool, not on
inference of causal models.  We will show how to interpret a
provenance graph as a causal model and relate syntactic and semantic
techniques for reasoning about causality in provenance graphs.

\section{Provenance Graphs}

For the purposes of this paper, we will employ a simplified model of
OPM-style provenance graphs, which we will just call provenance
graphs.  Fix a set of \emph{data values} $D$, and a set of
\emph{process names} $P$.  Recall that a bipartite graph $G = (V,W,E)$
is a (here, directed) graph with vertices $V \cup W$ and edges $E
\subseteq (V \times W )\cup( W \times V)$.  A \emph{provenance graph}
is a bipartite graph in which the vertices are labeled.  We call the
vertices of $V$ the \emph{artifacts} and the call vertices of $W$ the
\emph{processes}, and the artifact nodes are labeled with data values
and the process nodes are labeled with process names.  Intuitively, an
edge $(a,p) \in E$ means that $a$ was generated by process $p$, also
written $a \wgb p$, and an edge $(p,a) \in E$ means that $a$ was used
by process $p$, also written $p \used a$.

We call a provenance graph \emph{functional} if for each process node
$p \in W$ there is exactly one artifact $a \in V$ such that $(p,a) \in
E$.  We also call a provenance graph \emph{sorted} if there is a
function $ar : P \to \mathbb{N}$ such that each process node $x$ has
$ar(x)$ inputs (with in-edges labeled $1, \ldots n$).  Note that a
sorted functional provenance graph is essentially a first-order term
with sharing, i.e. a first-order term with nonrecursive let-binding.
However, for the purposes of this paper this sharing is important and
it is more convenient to think in terms of graphs.

In this paper, we consider the simple case where provenance graphs are
functional and both provenance graphs and causal graphs are acyclic.
(Note that acyclic causal models are called ``recursive'' in the
causal models literature.)

Now let us consider the provenance recording scenario described in the
introduction: say we have some program that computes a function $f :
D^n \to D$ whose implementation is unknown.  We assume a fixed set of
artifact nodes $V$ that includes at least $n$ input nodes
$v_1,\ldots,v_n$ and a distinguished result node $v_0$.  A
\emph{provenance graph semantics} for $f$ is a function
$\prov{f} : D^n \to PG(V)$, where $PG$ is the set of all provenance
graphs over $V$.

What properties should a provenance graph semantics have, beyond
simply assigning each input a graph?  We argue that there should be
some relationship between the function $f$ being described and the
``meaning'' of the provenance graph.  But for this to be sensible, we
first need to assign the graph a computational meaning.

Given a sorting $ar$, an interpretation $\SB{-}$ associates each
process name with a function $\SB{p} : D^{ar(p)} \to D$.  (Such an
interpretation is essentially an algebra over the first-order
signature $(P,ar)$.)  Given a graph $G$ with $n$ input nodes
$v_1,\ldots,v_n$, we can therefore define a function $\SB{G} : D^n \to
D$ that takes a tuple of values for the input nodes, applies the
interpretation functions to calculate the values for all of the nodes,
and finally returns the value of the result node $v_0$.

Given a provenance semantics $\prov{f}$ for an unknown function $f$,
we say that $\prov{f}$ is a \emph{pointwise approximation} of $f$ if
for each input $\vec{u}$,  we have $\SB{\prov{f}(\vec{u})} (\vec{u}) =
f(\vec{u})$.  In other words, we can recompute the value of $f$ at
$\vec{u}$ directly using the provenance graph returned for $\vec{u}$.

Most graph-based provenance semantics implicitly satisfy this
property.  But observe that there is always a trivial pointwise
provenance semantics defined by taking $\prov{f}(\vec{u})$ to be a
graph where the result node has no connection to the inputs and is
labeled with $f(\vec{u})$.  This is a pointwise approximation, but
gives us no insight into how the result depends on the inputs.  Thus,
capturing the original function pointwise is intuitively necessary,
but not sufficient for the provenance graph to be informative.
Something more is needed.

Given a provenance semantics $\prov{}$ for $f$, we say that $\prov{}$
approximates $f$ \emph{globally} provided that for each $\vec{u}$, we
have $\SB{\prov{f}(\vec{u})} = f$.  In other words, a global
approximation yields a provenance graph that simulates $f$ on all
inputs.  Note that a global approximation exists if (and only if) 
the language of provenance graphs is rich enough to express $f$.
Specifically, we define $\prov{f}$ as the constant function that returns a
provenance graph that simply say ``apply $f$''.  Of course, if $f$ is
a large, opaque program then this may not be helpful.

The global criterion seems very strong: it requires the
provenance graph to describe what would happen under an arbitrary
change to the input, which may be far more information than we need to
understand just one run of the program.  Moreover, a global
approximation semantics may not exist for reasonable combinations of
provenance graph languages and semantics.  For example, if the
provenance graphs can make use of only arithmetic functions then we
cannot globally approximate the behavior of a program that calculates
a properly recursive function such as factorial --- there is no fixed
arithmetic circuit that calculates factorials.  On the other hand, for
certain simple classes of workflows, global approximation is possible,
and clearly desirable.

Instead of an all-or-nothing correctness property, we now consider the
possibility of measuring how powerful a provenance semantics is as a
way of predicting the function it is meant to represent.

Let $\prov{}$ be a provenance semantics for $f$.  We define a relation
on input tuples $\vec{u}$ as follows:
\[\vec{u} \predicts \vec{u}' \iff \SB{\prov{f}(\vec{u})}(\vec{u}') = f(\vec{u}')\]
In other words, $\vec{u} \predicts \vec{u}'$ holds just in case the
provenance graph for $f$ on $\vec{u}$ correctly predicts the value of
$f$ on $\vec{u}'$.  That is, the relation $\predicts$ quantifies how
the generality of the provenance graphs produced by $f$.  We call the
relation $\predicts$ the \emph{predictive power} of provenance
semantics $P$.

Observe that $\predicts$ is reflexive if and only if $\prov{}$ is
pointwise approximation, and it is total if and only if $\prov{}$ is a
global approximation.  Furthermore, given two provenance semantics
$\prov[1]{}$ and $\prov[2]{}$, we can compare them by comparing their
predictive power $\predicts_1$ and $\predicts_2$.  Specifically, if
$\predicts_1 \subseteq \predicts_2$ then $\prov[2]{}$ has at least as
much predictive power as $\prov[1]{}$, which we might write formally
as $\prov[1]{} \leq \prov[2]{}$.  Note that two provenance semantics
might have equal predictive power but be very different as functions.

Given a language of provenance graphs and a fixed interpretation of
the process names they may use, consider the problem of designing a
provenance semantics for a given function $f$ whose predictive power
is as great as possible.  Some interesting questions we will leave for
future work include: Is there a unique ``most powerful'' semantics?
Given the program text of $f$, is there an effective way to produce
the most powerful (or a reasonably good) semantics?

\section{Causal models}

In the discussion so far, the function $f$ being investigated is
essentially a black box --- we can provide inputs and observe the
output, and nothing else.  There may be intermediate states and values
calculated by $f$ that are not visible externally, nor can we gain
understanding of $f$ by artificially altering intermediate states.
Understanding can often be aided by considering hypothetical changes
in the middle of a process.  For example, suppose $f$ is defined as
follows:
\[f(x,y,\vec{z}) = \frac{g(\vec{z})}{(x^2-1)y}\]
Then $f(1,0,\vec{z})$ is undefined (division by zero), but there is no
way to change just one input variable to a value such that $f$ is
defined.  Of course, inspecting the definition of the function makes
it obvious that both $x$ and $y$ need to change, but if $f$ is a black
box and $x$ and $y$ are just two of many input parameters then finding
a way to change the inputs to fix the problem would require many blind
guesses.  If we could see inside of $f$ and possibly change some
intermediate error results to dummy valid results, then we might be
able to find the problem more easily.

Structural causal models are (from a certain point of view) models of
computation that permit some inspection and intervention upon the
intermediate results.  We will now review standard definitions of
causal models and actual causes from
\cite{halpern05bjps-1,halpern05bjps-2}

A \emph{causal model} $M = (U,V,F)$ is a structure where $U$ is a set
of \emph{inputs} (or \emph{exogenous variables}), $V$ is a set of
 (\emph{endogenous}) variables, and $F$ is a family of
functions $F_X : (D^{V\cup U} \to D)$ mapping each vertex $X$ in $V$
to a function from tuples of data values to data values indexed by
$V$.  A \emph{causal graph} $G = (U \cup V, E)$ for $M$ is a directed
graph whose vertices are inputs or events of $M$ and such that $F_X$
depends only on the values corresponding to parent vertices of $X$ in
$G$.  We restrict attention to causal models whose underlying graph is
acyclic (these are sometimes called ``recursive'' models in the
structural-models literature).  For each acyclic causal model there is
a unique least causal graph expressing all and only the true
dependencies of the functions $F_X$.

A \emph{valuation} is simply a $V$-indexed tuple of values in $D$, or
$\sigma : D^V$.  We say a valuation is \emph{consistent} if for each
$X$ we have $F(X)(\sigma) = \sigma(X)$.  A causal model paired up with
a (consistent) valuation is called a (\emph{consistent}) \emph{causal
  situation}.

\emph{Interventions} are a distinctive feature of causal models.
Given a model $M$, we can form another models $M_{[X:=x]}$ by fixing
the value of vertex $X$ to $x$ and making the value of $X$ independent
of its former inputs.  This reflects the fact that although the causal
model represents the behavior of a closed system, we can reach into
the system (at least as a thought experiment) and change $X$ to a
value of our choice.

\begin{definition}
  If $M = (V,U,F)$ is a causal model, then the result of setting $X$
  to $x$ in $M$ is $M_{[X:=x]} = (V,U,F')$, where:
\begin{eqnarray*}
F'_Y(\sigma) &=& \left\{
  \begin{array}{ll}
     x &\mid X = Y\\
    F_Y(\sigma) &\mid X \neq Y
  \end{array}
\right.
\end{eqnarray*}
If the appropriate variable $X$ is clear from context, this may be written $M_x$.
\end{definition}

We also review Halpern and Pearl's definition of ``actual cause''.
\begin{definition}[Actual cause]
  Let $(M,\sigma)$ be a causal situation.  Let $\vec{X}$ be a subset
  of $V$ and $Y \in V$, and suppose $\vec{x} = \sigma(\vec{X})$ and $y
  = \sigma(Y)$.  Suppose that:
  \begin{enumerate}
  \item $\sigma(\vec{X}) = \vec{x}$ and $\sigma(Y) = y$.
  \item Some set of variables $W \subseteq V - X$ and values $\vec{x}'
    \in D$, and $\vec{w}'\in D$ exist such that:
    \begin{enumerate}
    \item $Y \neq y$ holds in $M_{\vec{x}',\vec{w}'}$
    \item $Y = y$ holds in $M_{\vec{x}, \vec{w}',\vec{z}}$ for all $Z \subseteq V
      - (X \cup W)$, where $\vec{z}$ are the values of $\vec{Z}$ in
      $M$.
    \end{enumerate}
  \end{enumerate}
  Then we say that $\vec{X} = \vec{x}$ is an \emph{weak cause} of $Y =
  y$.  Moreover, if no proper subset of $\vec{X} = \vec{x}$ is a weak
  cause, then $\vec{X} = \vec{x}$ is an \emph{actual cause} of $Y =
  y$.
\end{definition}

The definition of actual cause deserves explanation.  Briefly, the
idea is that an actual cause must first describe the true state of
affairs (part 1).  Secondly, there must be a way to change the value
of $Y$ by changing the values of $X$ (possibly plus some other
variables $W$), and there must not be a way of changing the value of
$Y$ by fixing the values of $X = \vec{x}$, $W = \vec{w}'$ and setting
any other variables to their original values.  (In fact, the variables
$Z$ are usually between $X$ and $Y$.)

For additional discussion we must refer the reader to Halpern and
Pearl~\cite{halpern05bjps-1}.  The discussion of actual
causes later in this paper does not depend heavily on the details of
this definition.

\section{Causal interpretations of provenance graphs}

Given a fixed interpretation of the process identifiers of a
provenance graph $G = (V,W,E)$, we can define a causal model and a
valuation $(M_G,\sigma_G)$. Suppose $V = U' \uplus V'$ where $U'$ is
the set of input nodes in $G$ (that is, artifact nodes not generated
by any process).  Let $M_G = (U',V' \cup W,F)$, where we define
$F_v(\sigma)$ for $v \in V'$ as the (unique) value of $\sigma(p)$ where
$p$ is the process node that generates $v'$, and $F_p(\sigma)$ is the
interpretation function for node $p$ in $G$, lifted to apply to
$\sigma$ by ignoring all values of nodes not used by $p$ in $\sigma$.
Moreover, we define the valuation $\sigma_G$ for $M_G$ by taking
$\sigma_G(v)$ to be the data value assigned to $v$ in $G$.

Just as the functional interpretation of a provenance graph can be
viewed as a local approximation of some unknown (functional) process,
its causal model interpretation can be viewed as an approximation of
some unknown (causal) process.  To make this idea precise, we
introduce an abstract notion of ``causal functions'' that support both
ordinary evaluation and intervention.

A \emph{causal function} over variables $(U,V)$ is a family of
functions $f_\tau : D^U \to D^V$, one for each partial valuation $\tau
: V \rightharpoonup D$.  We write $f$ for $f_{\emptyset}$, and we
require that for each partial valuation $\tau$ and input valuation
$\sigma$ we have $\tau \subseteq f_{\tau}(\sigma)$. Thus, for example,
$f(\sigma)$ computes all of the values of variables directly from the
inputs without any interference, while $f_{[X := x,Y := y]}(\sigma)$
calculates the values when $X$ is forced to be $x$ and $Y$ is forced
to be $y$. Note that a causal model $M$ defines a unique causal
function, which we will write $\SB{M}$.  On the other hand, the
definition of causal function makes sense without talking about the
specific propagation mechanism used to calculate the values.

Suppose that $f$ is a causal function with inputs $U$ and variables
$V$, and let $\prov{f}$ be a function mapping input values $\vec{u}
\in D^U$ to causal models $\prov{f}(\vec{u})$ over inputs $U$ and
variables $V$.  Then $\SB{\prov{f}(\vec{u})}$ is again a causal
function for each $\vec{u}$.

We say that a provenance semantics $\prov{}$ approximates $f$
\emph{pointwise} if for any $\vec{u}$, we have
$\SB{\prov{f}(\vec{u})}(\vec{u}) = f(\vec{u})$.  That is, both the end
results and intermediate states of $\prov{f}$ agree with $f$.  As with
pointwise approximations in the functional case, this is a rather weak
specification, since it is satisfied by defining $\prov{f}(\vec{u})$
as a disconnected causal process that simply defines each variable $X
\in V$ to be a constant $f(\vec{u})(X)$.

Likewise, we can define $\prov{f}$ to be a global approximation of $f$
when $\SB{\prov{f}(\vec{u})} = f$ for any $\vec{u}$.  Again, this is
very strong, since it says that $f$ must be expressible as a single
causal model --- that is, the provenance graph for $f$ tells us
everything about the causal structure of $f$.

Casual functions admit a third natural notion of approximation, which
is strictly in between pointwise and global approximation.  We say
that provenance semantics $\prov{f}$ approximates $f$ \emph{locally}
if for any $\vec{u}$, we have
\[\SB{\prov{f}(\vec{u})}_\tau(\vec{u}) = f_\tau(\vec{u})\]
That is, approximating $f$ locally does not require that showing how $f$
would behave on arbitrary inputs, but does ensure that we can predict
how the particular run of $f(\vec{u})$ would change if we had
intervened by fixing the values of the variables in $V$.

Local approximation is strictly stronger than pointwise approximation:
clearly local implies pointwise, but for a simple process such as $Y
:= X + 1; Z:= Y * 2$ the trivial pointwise approximation is not a
local approximation.  The requirement to handle arbitrary
interventions forces us to do more than simply record the actual
values of the intermediate variables; instead, we must record at least
some information about how they were related.

On the other hand, global approximation obviously implies local
approximation, but not the converse.  Local approximation is strictly
weaker than global approximation, since there is no requirement that
the causal model obtained for a run on input $\vec{u}$ will be useful
for predicting results or causal structure of $f$ running on another
input.  This allows different causal models to be used for different
inputs, as long as the causal structure of the model chosen for $f$
only depends on depend on the input parameters, not intermediate
variable values.

For example, consider $f(u;x,y) = (x+y)^u$, where the provenance graph
may only employ multiplication and addition.  Then there is no single
provenance graph that approximated $f$ globally, but for each choice
of $u$ there is a graph that describes the causal function $x,y
\mapsto f(u;x,y)$.

As with the functional case, we might also expect a causal model to
have partial explanatory power for other input settings.
Specifically, we define the predictive power of $\prov{f}$ as the
relation:
\[\vec{u} \predicts \vec{u}' \iff \forall
\tau. \SB{\prov{f}(\vec{u})}_\tau(\vec{u}') = f_\tau(\vec{u}') 
\]
In other words, $\vec{u} \predicts \vec{u}'$ holds when the causal
model generated from the provenance of $\vec{u}$ is a faithful model
of the behavior of $f$ at $\vec{u}'$ under arbitrary interventions.

For this definition of predictive power, local and global
approximation correspond to reflexivity and totality of $\predicts$.
As with the functional case, we can compare provenance semantics by
their predictive power.

\section{Inferences about causality in provenance}

The OPM standard also describes inferences on provenance graphs.
These inferences are formalized as Datalog-style inference rules that
allow us to infer new edges from existing edges.  For our provenance
graphs, the relevant rules from the OPM standard include:
\begin{eqnarray*}
  x \wdf y &\ent& x \wgb p \wedge p \used y\\
  p \wtb q  &\ent&p \used x \wedge x \wgb q\\
  x \wdf^+ y & \ent & x \wdf y \vee (x \wdf z \wedge z \wdf^+ y)\\
  p \wtb^+ q & \ent & p \wtb q \vee (p \wtb r \wedge r \wtb^+ q)
\end{eqnarray*}
For example, if process $p$ ``used'' artifact $x$ which ``was
generated by'' process $q$, then we infer that $p$ ``was triggered
by'' $q$.  However it is important to note that the terms ``used'',
``was generated by'' and ``influenced'' are simply different edge
labels whose relationship is being axiomatized by the Datalog rules
--- there is no connection in the standard to the causal-model
interpretation of provenance graphs which we have introduced in this
paper.

We want to give these edge relations meaning in terms of the causal
model interpretation, and use this as a basis for judging the
correctness of inferences.  In particular, we would like the edges to
correspond to causal relationships, as they are expected to do in OPM.

Consider a naive interpretation in which we interpret $p \used x$ as
meaning that $x = \sigma(x)$ was part of an actual cause of $p =
\sigma(p)$ in $M_G$, and likewise interpret $x \wgb p$ as meaning that
$p = \sigma(p)$ was part of a actual cause of $x= \sigma(x)$ in $M_G$.
However, this is a little too naive, since it is possible for $X = x$
to actually cause $Y = y$ and $Y = y$ to actually cause $Z= z$, while
in OPM graphs, the $\used$ and $\wgb$ edges should not have this
behavior.  We can fix this by further constraining these relations to
hold only when there is no other actual cause ``between'' $p$ and $x$.
Using these definitions of the basic edges, we can extrapolate
meanings for the other edges using the Datalog rules: for example, $x
\wdf y$ means that $Y = y$ is part of an actual cause of $X = x$ and
there is no other artifact strictly between $x$ and $y$, and $\wdf^+$
is the transitive closure, which (for a finite causal model) is
equivalent to saying that $Y = y$ is part of an actual cause of $X =
x$ without any constraints.  (We leave these observations as
conjectures for now.)

However, there is an important problem with the above situation:
namely, we have defined the meanings of the edges in terms of the
definition of actual cause, not in terms of the presence or absence of
edges in $M$.  Although every actual cause relationship corresponds to
an edge, there may also be edges that are syntactically present in
$M_G$ but do not correspond to immediate actual causes.  The possible
flows of information exhibited by the causal model do not always
correspond to actual causes.  In other words, the Datalog rules are
complete, but unsound, with respect to actual causality; the naive
approach of simply following all syntactic links to find all nodes
reachable from a given node is a safe over-approximation (but it may be
very inaccurate).

To clarify this point, we review some of the complexity results of
Eiter and Lukasiewicz~\cite{eiter02ai}.  For deterministic causal
models, they study the complexity of several definitions of causality,
including the ``actual cause'' definition introduced by Halpern and
Pearl~\cite{halpern05bjps-1}.  They show that determining whether a
set of variables and values is an actual cause of another event is
NP-hard, even for Boolean causal models.  These complexity lower
bounds hold whether or not the function definitions are considered
part of the problem, since in the Boolean case the amount of
information needed to describe the operators is small.  Datalog
queries can be computed in polynomial time~\cite{immerman86ic}, so
cannot in general correctly calculate actual causes.

\section{Related and future work}

As discussed in the introduction, we draw on previous work on
causality based on structural models in artificial intelligence (see
for example ~\cite{pearl00causal} for a comprehensive discussion of
this work and its applications).  The worst-case complexity and
special cases of computing actual causes and causal explanations is
comprehensively studied by a series of papers by Eiter and
Lucasiewicz~\cite{eiter02ai,eiter04ai,eiter06ai}.  As we have noted,
these have immediate applications to showing that simple transitive
closure algorithms for extracting actual causes from provenance cannot
suffice for provenance inference, at least if Halpern and Pearl's
definitions are used.

Halpern gave a well-received invited presentation on causality,
responsibility and blame at a recent workshop on
provenance~\cite{cheney09tapp-sigmod}.  The analogy between causal
models and provenance has been ``in the air'' since then, and has been
discussed in more recent work by Chapman et al.~\cite{chapman10tapp}
and an overview paper by Cheney et al.~\cite{cheney09onward}

The treatment of causality here plays a similar role to dependence in
previous work with Acar and Ahmed on dependency
provenance~\cite{cheney07dbpl}, and subsequently on a ``complete
provenance trace'' model~\cite{traces-tr}.  These models support some
inference concerning how hypothetical changes to (parts of) the input
affect (parts of) the output.  However, these techniques are based on
a database query language supporting arbitrarily nested sets and
records, which do not seem to match structural causal models over
atomic data.  Acar et al.~\cite{acar10tapp} presents a translation
from a core database query language to OPM-style provenance graphs;
the graphs in that work do not have a causal or computational
interpretation, but exploring this possibility is an obvious next
step.

In this paper, we have focused on a (perhaps deceptively) simple
setting where the provenance graphs are directly analogous to causal
models.  It is not clear how far this analogy can be pushed using
known forms of causal models; it is possible that extensions to the
causal framework are needed to provide semantics for provenance models
describing richer computational models.  For example, matters likely
become more interesting in the presence of concurrent processes that
may interact with one another (see for example~\cite{sassone09tapp}).
This is a common case in scientific workflow provenance and causal
provenance semantics should be developed for (or adapted to) models
such as dataflow networks or process.

There are several other developments in structural causal models that
would be interesting to explore from the point of view of provenance,
including using probabilistic causal models to explain how provenance
can make unreliable computations statistically ``repeatable'', and
using existing definitions of explanation, responsibility and blame
for provenance.

Finally, as illustrated by the ``cake'' example, provenance records
sometimes employ an implicit notion of state and resource consumption.
For example, in baking a cake, some inputs are ``consumed'' (e.g. the
ingredients), while others are only catalysts (e.g. the cake pan).
The provenance model also does not reflect time or resource usage.  Of
course, this kind of information can be added to the data model (and
the OPM draft already allows for both time and domain-specific
annotations), but it would be interesting to explore applications of
linear or resource-sensitive logics to this setting.

\section{Conclusions}

We have explored an analogy between the OPM-style provenance graphs
that are becoming popular in scientific workflow systems (and other
settings) and structural causal models as developed by Pearl, Halpern
and others in artificial intelligence.  In particular, a provenance
graph can be interpreted as a causal model in a natural way, provided
that we know how to interpret the processing steps as functions. We
considered the problem of using provenance graphs to represent
particular runs of a program.  Existing definitions and results
concerning causal models can be lifted to provenance graphs.

 We also investigated the problem of
inferring actual causes over provenance graphs using Datalog-style
rules.  As shown by the complexity of computing or calculating
explanations in the Halpern-Pearl approach is quite high (usually at
least NP-hard), which shows that, for example, queries about actual
cause or explanation cannot be expressed using Datalog rules, even in
the special case of provenance graphs over Boolean functions.  This
shows that Halpern and Pearl's definitions of actual cause and the ad
hoc rules that have been proposed for reasoning about provenance
graphs are incompatible.

These observations are not deep --- they amount to applying known
results in the structural causality literature to analyze informal
claims in the provenance literature.  Adopting a different
model of causality might lead to different conclusions.  Nevertheless,
we believe that the exercise of trying to formalize the relationship
between provenance and causal explanation has not been seriously tried
before and is worthwhile, if for no other reason than to provoke
others to do a better job.

\bibliographystyle{eptcs} 
\bibliography{paper}
\end{document}